\documentclass[final]{siamltex}

\usepackage{graphicx}
\usepackage{subfigure}
\usepackage[hyphens]{url}
\usepackage[colorlinks,unicode,linkcolor=black]{hyperref}
\usepackage{xspace}
\usepackage{listings}
\newcommand{\code}[1]{\lstinline$#1$}
\newcommand{\figref}[1]{Fig.~\ref{#1}}
\newcommand{\eqref}[1]{(\ref{#1})}

\newcommand {\de} {\mbox{d}}

\newcommand {\rem}[1]{}

\newcommand{\addpp}[1]{{#1\nolinebreak[4]\hspace{-.05em}\raisebox{.4ex}{\tiny\bf ++}}\xspace}
\newcommand{\Cpp}{\addpp{C}}

\title{Programming CUDA and OpenCL:\\a Case Study Using Modern C++ Libraries}

\author{
Denis Demidov\thanks{
Kazan Branch of Joint Supercomputer Center,
Russian Academy of Sciences,
Lobachevsky st. 2/31, 420111 Kazan, Russia
({\tt ddemidov@ksu.ru}) }
\and Karsten Ahnert\thanks{Ambrosys GmbH, Geschwister-Scholl-Stra\ss e 63a, 14471 Potsdam, Germany ({\tt karsten.ahnert@gmx.de}) }
\and Karl Rupp\thanks{Mathematics and Computer Science Division,
Argonne National Laboratory,
9700 South Cass Avenue, Argonne, IL 60439, USA
({\tt rupp@mcs.anl.gov}) }
\and Peter Gottschling\thanks{SimuNova, Helmholtz-Str. 10, 01069 Dresden \&
Inst. Scientific Computing, TU Dresden, 01062 Dresden}
({\tt peter.gottschling@simunova.com}) }

\rem{Institut f\"ur Physik und Astronomie, Universit\"at Potsdam, Karl-Liebknecht-Strasse 24/25, 14476 Potsdam-Golm, Germany}

\begin{document}

\maketitle

\begin{abstract}
    We present a comparison of several modern \Cpp libraries providing high-level interfaces
    for programming multi- and many-core architectures on top of CUDA or OpenCL.
    The comparison focuses on the solution of ordinary differential equations and is based on odeint,
    a framework for the solution of systems of ordinary differential equations. Odeint is designed in a
    very flexible way and may be easily adapted for effective use of libraries such
    as MTL4, VexCL, or ViennaCL, using CUDA or OpenCL technologies.
    We found that CUDA and OpenCL work equally well for problems
    of large sizes, while OpenCL has higher overhead for smaller problems.
    Furthermore, we show that modern high-level libraries allow to effectively
    use the computational resources of many-core GPUs or multi-core CPUs without much
    knowledge of the underlying technologies.
\end{abstract}



%
% INTRODUCTION
%
\section{Introduction}

\pagestyle{myheadings}

\thispagestyle{plain}
\markboth{D.~DEMIDOV, K.~AHNERT, K.~RUPP, AND P.~GOTTSCHLING}{PROGRAMMING CUDA AND OPENCL\ldots}

Recently, general purpose computing on graphics processing units (GPGPU) has
acquired considerable momentum in the scientific community. This is confirmed
both by increasing numbers of GPGPU-related publications and GPU-based
supercomputers in the TOP500\footnote{ \href{ http://top500.org }{
http://top500.org }} list. Major programming frameworks are NVIDIA CUDA and
OpenCL.
The former is a proprietary parallel computing architecture developed
by NVIDIA for general purpose computing on NVIDIA graphics adapters, and the
latter is an open, royalty-free standard for cross-platform, parallel
programming of modern processors and GPUs maintained by the Khronos group. By
nature, the two frameworks have their distinctive pros and cons. CUDA has a
more mature programming environment with a larger set of scientific libraries,
but is available for NVIDIA hardware only. OpenCL is supported on a wide range
of hardware, but its native API requires a much larger amount of boilerplate
code from the developer. Another problem with OpenCL is that it is generally
difficult to achieve performance portability across different hardware
architectures.

Both technologies are able to provide scientists with the vast computational
resources of modern GPUs at the price of a steep learning curve.  Programmers
need to familiarize themselves with a new programming language and, more
importantly, with a new programming paradigm. However, the entry barrier may be
lowered with the help of specialized libraries. The CUDA Toolkit includes
several such libraries (BLAS implementations, Fast Fourier Transform, Thrust
and others). OpenCL lacks standard libraries, but there are a number of
third-party projects aimed at developing both CUDA and OpenCL programs.

% Using such libraries it is also possible to write cross-platform
% code. For example, VexCL executables can be configured by an
% environment variable to run on either the GPU or the CPU. Thrust supports
% several computation backends, such as CUDA, OpenMP, or Intel Thread Building
% Blocks. One of the backends may be selected during the compilation stage by a
% compiler option.

% [DD]: I don't like this paragraph. Hardware portability is intrinsic feature
% of any OpenCL application, not just VexCL, and one does not have to use
% libraries for that.

This paper presents a comparison of several modern \Cpp libraries aimed at ease
of GPGPU development. We look at both convenience and performance of the
libraries under consideration in the context of solving ordinary differential
equations.  The comparison is based on odeint\footnote{\href{
http://odeint.com }{ http://odeint.com } }, a modern \Cpp library for solving
ordinary differential equations (ODEs) numerically  \cite{OdeintRef2,OdeintRef1}
which has been included into the Boost
libraries\footnote{ \href{ http://boost.org } { http://boost.org } } recently.
It is developed in a generic way using template meta-programming techniques,
which leads to extraordinary high flexibility at utmost performance. The
numerical algorithms are implemented independently of the underlying
arithmetics. This results in a broad applicability of the library, especially
in non-standard environments.  For example, odeint supports matrix types,
arbitrary precision arithmetics, and can be easily adapted to use either CUDA
or OpenCL frameworks.

The GPGPU libraries considered in this work are MTL4, VexCL, and ViennaCL. We
also employ Thrust\footnote{ \href{ http://thrust.github.com }{
http://thrust.github.com }} in order to provide a reference point for the
comparison of the considered libraries.  Thrust is a parallel algorithms
library which resembles the \Cpp Standard Template Library \cite{ThrustRef}.
Its high-level interface greatly enhances developer productivity while enabling
performance portability between GPUs and multi-core CPUs.  Thrust is
distributed with the NVIDIA CUDA Toolkit since version~4.1.

\begin{description}
    \item[MTL4] (The Matrix Template Library)\footnote{ \href{ http://mtl4.org }{
        http://mtl4.org }} is a \Cpp linear algebra library providing
        an intuitive interface by establishing a domain-specific language
        embedded in \Cpp~\cite{pg_ccgrid12}.
        The library aims for maximal performance achievable by high-level languages
        using compile-time transformations.
        % For this purpose, source code is transformed during compile time by expression
        % templates and meta-tuning.
        Currently, three versions exist:
        the open-source edition supporting single- and multi-core
        CPUs, the supercomputing edition providing generic MPI-based
        parallelism, and the CUDA edition introduced in this
        paper. In the following, we will refer to the CUDA version of
        MTL4 as CMTL4.
    \item[VexCL] is a vector expression template
        library\footnote{ \href{ https://github.com/ddemidov/vexcl }{
        https://github.com/ddemidov/vexcl }} for OpenCL \cite{VexCLRef}. It has
        been created for ease of OpenCL development with \Cpp.  VexCL strives to
        reduce the amount of boilerplate code needed to develop OpenCL
        applications. The library provides a convenient and intuitive notation
        for vector arithmetic, reduction, sparse matrix-vector multiplication,
        etc.  Multi-device and even multi-platform computations are supported.
    \item[ViennaCL] (The Vienna Computing Library) is a scientific computing
        library\footnote{ \href{ http://viennacl.sourceforge.net }{
        http://viennacl.sourceforge.net }} written in \Cpp \cite{ViennaCLRef}.
        CUDA and OpenMP compute backends were added recently, but
        only the initial OpenCL backend is considered in the remainder of this work.
        The programming interface is compatible with
        Boost.uBLAS\footnote{ \href{ http://www.boost.org/libs/numeric/ublas }
        { http://www.boost.org/libs/numeric/ublas } }
        and allows for simple, high-level access to the vast
        computing resources available on parallel architectures such as GPUs.
        The library's primary focus is on common linear algebra operations (BLAS
        levels 1, 2 and 3) and the solution of large sparse systems of equations by
        means of iterative methods with optional preconditioners.
\end{description}

CUDA and OpenCL differ in their handling of compute kernels compilation. In
NVIDIA's framework the compute kernels are compiled to PTX code together with
the host program. PTX is a pseudo-assembler language which is compiled at
runtime for the specific NVIDIA device the kernel is launched on. Since PTX is
already very low-level, this just-in-time kernel compilation has low overhead.
In OpenCL the compute kernels are compiled at runtime from higher-level C-like
sources, adding an overhead which is particularly noticeable for smaller
sized problems. A portable pre-compilation to some low-level pseudo-code as in CUDA is
not feasible in OpenCL because of hardware agnosticism by design.

The approach taken for the generation and compilation of the compute kernels is
one of the main differences between the OpenCL libraries we considered.
VexCL generates and compiles an OpenCL
program with a single kernel for each vector expression it encounters.  This
leads to potentially higher initialization overhead, but should prove to be
more effective in long runs. On the other hand,
ViennaCL uses a set of predefined kernels, which functionally overlaps with
BLAS level 1 routines for vector operations. These kernels are compiled in
batch at the program start to allow for faster initialization. However, due to
this design decision, vector expressions with several operands may result in the launch
of more than one kernel. It should be noted that because of the main focus
of ViennaCL being on iterative solvers for large sparse systems of equations,
where complex vector expressions are rare, predefined kernels are favorable in
such a setting.  
%The implications of these two distinct design decisions in
%VexCL and ViennaCL will be further addressed in the discussion of the results. %[KR]: This is not further addressed actually...

The other difference between CUDA and OpenCL is that CUDA supports a subset of
the \Cpp language in compute kernels, while OpenCL kernels are written in a subset
of C99. Therefore, CUDA programmers may use template meta-programming techniques
which may lead to more efficient and compact code. The native OpenCL API does
not provide such features, but the drawback is balanced by the ability of
kernel source generation during runtime. Modern \Cpp libraries such as those
considered in this work successfully use this approach and hide low-level details
from their users.

%
% ADAPTING ODEINT
%
\section{Adapting odeint} \label{sec:adapting-odeint}

Ordinary differential equations play a major role in many scientific
disciplines. They occur naturally in the context of mechanical systems, like
granular \cite{poschel_computational_2005} and molecular dynamics. In fact, the
Newtonian and Hamiltonian mechanics are formulated as ODEs
\cite{landau_mechanics_1976}.  Many other applications can be found in such
diverse fields as biology \cite{brauer_mathematical_2001,Murray-93},
neuroscience \cite{izhikevich_dynamical_2006}, chemistry
\cite{atkins_physical_2001}, and social sciences \cite{Helbing01}. Furthermore,
ODEs are also encountered in the context of the numerical solution of
non-stationary partial differential equations (PDEs), where they occur after
a discretization of the spatial coordinates \cite{Hundsdorfer2003}.

Odeint solves the initial value problem (IVP) of ordinary differential
equations given by
\begin{equation}
\frac{\de x}{\de t } = \dot{x} = f(x , t), \quad \quad x(0) = x_0.
\label{eq:ode}
\end{equation}
Here, $x$ is the dependent variable and is usually a vector of real or complex
values.  $t$~is the independent variable. We will refer to $t$ as the time
throughout the article and denote the time derivative with $\de x / \de t =
\dot{x}$. $f(x,t)$ is the system function and defines the ODE.

Typical use cases for solving ODEs on GPUs are large systems of coupled ODEs
which occur as discretizations of PDEs, or ODEs defined on lattices or
graphs. Another use case are parameter studies, where the
dependence of an ODE on some parameters is of interest. Here, a high-dimensional ODE consisting of many low-dimensional
uncoupled ODEs, each with a different parameter set, is
considered. This one large system is then solved at once, hence all
low-dimensional ODEs are solved simultaneously.

Numerous methods for solving ODEs exist \cite{HairerSolvingODEI,
HairerSolvingODEII,Press-92}, which are usually categorized in the field of numerical
analysis.  Odeint implements the most prominent of these methods, for example
the classical Runge-Kutta methods and Runge-Kutta-Fehlberg methods, multi-step
methods (Adams-Bashforth-Moulton), symplectic Runge-Kutta-Nystr\"om methods,
and implicit methods (Rosenbrock and implicit Euler). All of these methods work
iteratively, starting from a given initial value $x(t_0)$ to calculate the next
value $x(t+\Delta t)$.  $\Delta t$ is the step size and may be chosen either
statically or adaptively.  For reference, we note that the simplest method is
the explicit Euler scheme
\begin{equation}
x\left(t+\Delta t\right) = x(t) + \Delta t \; f(x(t),t) .
\label{eq:euler}
\end{equation}
Its global accuracy is of first order, but the scheme is usually not
used for real applications because of stability and accuracy issues.

One main feature of odeint is the decoupling of the specific algorithm
for solving the ODE from the underlying arithmetic operations. This
is achieved by a combination of a state type, an algebra, and
operations. The state type represents the state of the ODE being
solved and is usually a vector type like \code{std::vector<>},
\code{std::array<>}, or a vector residing on a GPU. The algebra is
responsible for iterating through all elements of the state, whereas
the operations are responsible for the elementary operations.

To see how the explicit Euler method \eqref{eq:euler} is translated to
code in odeint, we briefly discuss its implementation:
\begin{lstlisting}
template< class State, class Algebra, class Operations >
class euler {
    // ...
    template< class Ode >
    void do_step(Ode ode, State &x, time_type t, time_type dt) {
        ode(x, m_dxdt, dt);
        Algebra::for_each3( x, x, m_dxdt, Operations::scale_sum2(1.0, dt) );
    }
};
\end{lstlisting}
The state type, the algebra, and the operations enter the Euler method
as template parameters, hence they are exchangeable. The function
object \code{ode} represents the ODE and must be provided by the
user. It calculates the right hand side $f(x,t)$ of \eqref{eq:ode} and
stores the result in \code{m_dxdt}.  The call of \code{for_each3}
iterates simultaneously over all elements of three vectors and applies
\code{scale_sum2} to each triple. The operation is performed in-place,
meaning that $x$ is updated to the new value.  In the code-snippet
above, the call to \code{for_each3} is thus equivalent to the vector
operation
\begin{lstlisting}
 x = 1.0 * x + dt * m_dxdt
\end{lstlisting}
which is just Eq.~\eqref{eq:euler}, since \code{m_dxdt} holds the values of
$f(x(t), t)$.

An odeint algebra is a class consisting of
\code{for_each1}, \dots, \code{for_eachN} methods. For example, the
\code{for_each3} method in the \code{range_algebra} --- the default algebra for
most vector types --- is similar to
\begin{lstlisting}
struct range_algebra {
    // ...
    template< class R1, class R2, class R3, class Op >
    void for_each3(R1 &r1, R2 &r2, R3 &r3, Op op) {
        auto it1 = boost::begin(r1);
        auto it2 = boost::begin(r2);
        auto it3 = boost::begin(r3);
        while( it1 != boost::end(r1) ) op(*it1++, *it2++; *it3++);
    }
    // ...
};
\end{lstlisting}
The operations are represented by a struct with public member classes defining the
operations used by the algebras. There is only one default
operations class implementation in odeint, which uses the arithmetic operators as usual:
\begin{lstlisting}
struct default_operations {
    // ...
    template< class Fac1, class Fac2 >
    struct scale_sum2 {
        Fac1 m_fac1;
        Fac2 m_fac2;
        scale_sum2( Fac1 fac1, Fac2 fac2 ) : m_fac1(fac1), m_fac2(fac2) { }
        template< class S1, class S2, class S3 >
        void operator()( S1 &s1, const S2 &s2, const S3 &s3 ) const {
            s1 = m_fac1 * s2 + m_fac2 * s3;
        }
    };
    // ...
};
\end{lstlisting}

The main reason for the separation of algebra and operations is that
all arithmetic calculations and iterations are completely encapsulated
into the algebra and the operations. Therefore, the numerical
algorithms to solve the ODEs are independent from the underlying
arithmetics. Note that the algebra and the operations must be chosen
such that they interact correctly with the state type.

Many libraries for vector and matrix types provide expression templates
\cite{Vandevoorde:CppTemplates,Veldhuizen:ExpressionTemplates, Veldhuizen:Techniques}
for the elementary operations using operator overload
convenience.  Such libraries do not need to define their own algebra, but can
instead be used with a default algebra and a default operation set included in
odeint, which simply call the operations directly on the matrix or vector type.

We describe the adaptation of odeint for the GPGPU libraries under
consideration in the following. The adaptations are now part of odeint, thus
native support for these libraries is available.  Implementation details such as
the resizing of vectors is accomplished in a straight-forward manner and not
further addressed for the sake of conciseness.

To adapt Thrust to odeint, we need to provide both an algebra and
operations. The algebra needs to define the \code{for_each} family of
algorithms. All of these operations follow the same pattern,
so we consider \code{for_each3} only:
\begin{lstlisting}
struct thrust_algebra {
    template<class StateType1, class StateType2, class StateType3, class Op>
    static void for_each3(StateType1 &s1, StateType2 &s2, StateType3 &s3, Op op) {
        thrust::for_each(
                thrust::make_zip_iterator( thrust::make_tuple(
                    s1.begin(), s2.begin(), s3.begin() ) ),
                thrust::make_zip_iterator( thrust::make_tuple(
                    s1.end(), s2.end(), s3.end() ) ),
                op);
    }
};
\end{lstlisting}
Here, \code{thrust::make_zip_iterator} is used in combination with
\code{make_tuple} to pack several device vector iterators into a single
iterable sequence.  The sequence is then processed by the
\code{thrust::for_each} algorithm, applying the function object \code{op} to
each entry.

The operations called via the function object \code{op} are defined in
\code{thrust_operations} and are actually function objects executed on the
respective CUDA device:
\begin{lstlisting}
struct thrust_operations {
    template<class Fac1 = double, class Fac2 = Fac1>
    struct scale_sum2 {
        const Fac1 m_alpha1;
        const Fac2 m_alpha2;

        scale_sum2(const Fac1 alpha1, const Fac2 alpha2)
            : m_alpha1(alpha1), m_alpha2(alpha2) { }

        template< class Tuple >
        __host__ __device__ void operator()( Tuple t ) const {
            thrust::get<0>(t) = m_alpha1 * thrust::get<1>(t)
                               + m_alpha2 * thrust::get<2>(t);
        }
    };
};
\end{lstlisting}
The device function object uses \code{thrust::get<>} functions to unpack the
zip iterator into separate values.  This approach is heavily used with Thrust
and allows to process several vectors in a single efficient sweep.

\begin{sloppypar}
  CMTL4, VexCL, and ViennaCL libraries provide convenient
  expression templates that may be directly used with odeint's
  \code{vector_space_algebra} and \code{default_operations}. This
  combination proved to be effective with CMTL4 and VexCL, where each
  expression results in a single kernel. For ViennaCL, however,
  default operations involving more than two terms result in multiple
  kernel launches.  Moreover, temporary vectors are allocated and
  deallocated for each of such composite operations, resulting in a
  dramatic decrease of performance.  To address such problems,
  ViennaCL provides a kernel generator \cite{tillet:kernel-generator},
  which is able to generate specialized operations for ViennaCL. 
  %These
  %use a custom kernel generation mechanism provided by the library. %[KR] Sentence is redundant
  For example, the \code{scale_sum2} operation is defined as:
\begin{lstlisting}
struct viennacl_operations {
    template<class Fac1 = double, class Fac2 = Fac1>
    struct scale_sum2 {
        // ...
        template<class T1, class T2, class T3>
        void operator()( viennacl::vector<T1> &v1,
                    const viennacl::vector<T2> &v2,
                    const viennacl::vector<T3> &v3) const
        {
            typedef viennacl::generator::vector<T1> vec;

            viennacl::generator::custom_operation op;
            op.add( vec(v1) = m_alpha1 * vec(v2) + m_alpha2 * vec(v3) );
            op.execute();
        }
    };
};
\end{lstlisting}
Here, a custom OpenCL kernel is automatically generated from symbolic vector
expression in the first call of the \code{operator()} and then buffered and reused for all
subsequent calls. The objects of type \code{vec} are used to distinguish direct ViennaCL statements
from symbolic specifications for the kernel generation facility.
\end{sloppypar}

%
% NUMERICAL EXPERIMENTS
%
\section{Numerical Experiments}

As shown in the previous section, all four GP\-GPU libraries considered
in our comparison could be adapted to odeint without getting in
contact with low-level CUDA or OpenCL code.  The purpose of this
section is to evaluate the performance of the GPGPU libraries and
whether there is a price to pay for the high-level interface.

%
% LORENZ ATTRACTOR
%

\subsection{Lorenz Attractor Ensemble}

In the first example we consider the Lorenz system \cite{Lorenz-63}. The
Lorenz system is a system of three coupled ODEs which shows chaotic
behavior for a large range of parameters. It is one of the most frequently
used ODEs for evaluation purposes in the nonlinear dynamics community.
The equations for the Lorenz system read
\begin{equation}
    \dot{x} = -\sigma \left( x - y \right), \quad
    \dot{y} = R x - y - xz, \quad
    \dot{z} = -bz + xy.
    \label{eq:lorenz}
\end{equation}

Solutions of the Lorenz system usually furnish very interesting
behavior in dependence on one of its parameters.  For example, one
might want to study the chaoticity in dependence on the parameter
$R$. Therefore, one would create a large set of Lorenz systems (each
with a different parameter $R$), pack them all into one system and
solve them simultaneously. In a real study of chaoticity one
may also calculate the Lyapunov exponents \cite{Ott-book-02}, which
requires to solve the Lorenz system and their linear perturbations.

The Thrust version of the system function object for the Lorenz attractor ensemble
example is presented below. It holds the model parameters and provides the necessary
\code{operator()} with a signature required by the odeint library. The state
type is represented by \code{thrust::device_vector<double>}:
\begin{lstlisting}
typedef thrust::device_vector<double> state_type;

struct lorenz_system {
    size_t N;
    const state_type &R;
    lorenz_system(size_t n, const state_type &r) : N(n), R(r) { }
    void operator()(const state_type &x, state_type &dxdt, double t) const;
};
\end{lstlisting}
The $X$, $Y$, and $Z$ components of the state are held in the continuous
partitions of the vector.  \code{operator()} uses the standard technique of
packing the state components into a zip iterator and passes the composite
sequence to the \code{thrust::for_each} algorithm together with the provided
device function object:
\begin{lstlisting}[firstnumber=12]
struct lorenz_functor;

void lorenz_system::operator()(const state_type &x, state_type &dxdt, double t) const
{
        thrust::for_each(
                thrust::make_zip_iterator( thrust::make_tuple(
                        R.begin(),
                        x.begin(), x.begin() + N, x.begin() + 2 * N,
                        dxdt.begin(), dxdt.begin() + N, dxdt.begin() + 2 * N ) ),
                thrust::make_zip_iterator( thrust::make_tuple(
                        R.end(),
                        x.begin() + N, x.begin() + 2 * N, x.end(),
                        dxdt.begin() + N, dxdt.begin() + 2 * N, dxdt.end() ) ),
                lorenz_functor() );
}
\end{lstlisting}
The device function object unpacks the individual components and applies the required
operations to the derivative part, essentially leading to a one-to-one
translation of \eqref{eq:lorenz} into code:
\begin{lstlisting}[firstnumber=last]
struct lorenz_functor {
    template< class T >
    __host__ __device__ void operator()( T t ) const {
        double R = thrust::get<0>(t);
        double x = thrust::get<1>(t);
        double y = thrust::get<2>(t);
        double z = thrust::get<3>(t);
        thrust::get<4>(t) = sigma * ( y - x );
        thrust::get<5>(t) = R * x - y - x * z;
        thrust::get<6>(t) = -b * z + x * y ;
    }
};
\end{lstlisting}

The system function object for the CMTL4 version of the Lorenz attractor example is
more compact than the Thrust variant because CMTL4 supports a rich set of vector
expressions.  CMTL4 provides the type \code{multi_vector} that allows for
expressing the operations directly:
\begin{lstlisting}
typedef mtl::dense_vector<double>      vector_type;
typedef mtl::multi_vector<vector_type> state_type;

struct lorenz_system {
    const vector_type &R;
    explicit lorenz_system(const vector_type &R) : R(R) { }

    void operator()(const state_type& x, state_type& dxdt, double t) {
	dxdt.at(0) = sigma * (x.at(1) - x.at(0));
	dxdt.at(1) = R * x.at(0) - x.at(1) - x.at(0) * x.at(2);
	dxdt.at(2) = x.at(0) * x.at(1) - b * x.at(2);
    }
};
\end{lstlisting}
In this context, the class \code{multi_vector} is used in two ways:
expressing  operations on sub-vectors and on entire vectors.
Each operation on sub-vectors of  \code{x} and
\code{dxdt} causes a kernel call.

There is potential for optimization when the three operations are performed by one
kernel.
This can be achieved with the following formulation:
\begin{lstlisting}[firstnumber=9]
    ( lazy(dxdt.at(0)) = sigma * (x.at(1) - x.at(0)) ) ||
    ( lazy(dxdt.at(1)) = R * x.at(0) - x.at(1) - x.at(0) * x.at(2) ) ||
    ( lazy(dxdt.at(2)) = x.at(0) * x.at(1) - b * x.at(2) );
\end{lstlisting}
The vector assignments are not performed immediately but their evaluation is delayed
and can be fused with other expressions~--- denoted by operator \code{||}. 
This formulation has yet another advantage: the three vector operations
are performed in one single loop which provides much better data locality
for vector \code{x}.
For performance sake, the multi-vectors are constructed with contiguous memory
whenever the types allow for it.
Then, expressions on multi-vectors can be evaluated with one kernel call.
Especially for small vectors, the overhead of calling multiple kernels
is significant: we observed 150\,\% overhead with 3-component vector with 4K
entries compared to one vector of size 12K.

The VexCL implementation of the Lorenz attractor ensemble example is as
compact as that of CMTL4. Here, the state is represented by the
\code{vex::multivector<double,3>} type, which holds three instances of
\code{vex::vector<double>} and transparently dispatches all operations to the
underlying components. The code for the body of \code{operator()} practically
coincides with the problem statement \eqref{eq:lorenz}:
\begin{lstlisting}
typedef vex::multivector<double, 3> state_type;

struct lorenz_system {
    const vex::vector<double> &R;
    lorenz_system(const vex::vector<double> &r) : R(r) {}

    void operator()(const state_type &x, state_type &dxdt, double t) const {
        dxdt(0) = sigma * (x(1) - x(0));
        dxdt(1) = R * x(0) - x(1) - x(0) * x(2);
        dxdt(2) = x(0) * x(1) - b * x(2);
    }
};
\end{lstlisting}

However, the drawback of this variant is that it leads to three kernel
launches, namely one per each vector assignment. As we have discussed previously
for the CMTL4 variant, this results in suboptimal
performance. A direct use of arithmetic operations for multi-vectors is
not possible due to mixed components in the right hand side
expressions.  These additional kernel launches can be eliminated in
VexCL by assigning a tuple of expressions to a multi-vector. The
required implementation is only slightly less intuitive than the above
variant:
\begin{lstlisting}[firstnumber=9]
    dxdt = std::tie(   sigma * (x(1) - x(0)),
                       R * x(0) - x(1) - x(0) * x(2),
                       x(0) * x(1) - b * x(2)          );
\end{lstlisting}
The performance gain of these fused expressions is a bit larger
(25\,\% for large systems) compared to CMTL4. 
The reason might be the larger kernel launch overhead for OpenCL
kernels.

For the ViennaCL version of the Lorenz attractor example a
\code{boost::fusion::vector} is used to pack the coordinate components of the
state vector into a single type. Individual components are instances of the
\code{viennacl::vector<double>} type.  The ViennaCL kernel generation facility already used in
Sec.~\ref{sec:adapting-odeint} is then used to avoid multiple kernel launches.
Even though a \code{custom_operation} object is instantiated in each call to \code{operator()},
the kernel is created only once and then buffered internally for further reuse.
\begin{lstlisting}
typedef fusion::vector<
    viennacl::vector<double>, viennacl::vector<double>, viennacl::vector<double>
    > state_type;

struct lorenz_system {
    const viennacl::vector<double> &R;
    lorenz_system(const viennacl::vector<double> &r) : R(r) {}

    void operator()(const state_type &x, state_type &dxdt, double t) const {
        typedef viennacl::generator::vector<value_type> vec;

        const auto &X = fusion::at_c<0>(x);
        const auto &Y = fusion::at_c<1>(x);
        const auto &Z = fusion::at_c<2>(x);

        auto &dX = fusion::at_c<0>(dxdt);
        auto &dY = fusion::at_c<1>(dxdt);
        auto &dZ = fusion::at_c<2>(dxdt);

        viennacl::generator::custom_operation op;
        op.add( vec(dX) = sigma * (vec(Y) - vec(X)) );
        op.add( vec(dY) = element_prod(vec(R), vec(X)) - vec(Y)
                        - element_prod(vec(X), vec(Z)) );
        op.add( vec(dZ) = element_prod(vec(X), vec(Y)) - b * vec(Z) );
        op.excecute()
    }
};
\end{lstlisting}

%
% PHASE OSCILLATORS
%
\subsection{Chain of Coupled Phase Oscillators}

As a second example we consider a chain of coupled phase
oscillators. A phase oscillator describes the dynamics of an
autonomous oscillator \cite{PhaseOscillator}. Its evolution is
governed by the phase $\varphi$, which is a $2\pi$-periodic variable growing linearly
in time, i.e.~$\dot{\varphi} = \omega$, where $\omega$ is the phase
velocity. The amplitude of the oscillator does not occur in this
equation, so interesting behavior can only be observed if many
of such oscillators are coupled. In fact, such a system can be used to
study such divergent phenomena as synchronization, wave and pattern
formation, phase chaos, or oscillation death
\cite{Kuramoto-84,Synchronization-Pikovsky}. It is a prominent example
of an emergent system where the coupled system shows a more complex
behavior than its constitutes.

The concrete example we analyze here is a chain of nearest-neighbor
coupled phase oscillators \cite{Cohen-Rand-Holmes-82}:
\begin{equation} \label{eq:phasesystem}
    \dot{\varphi}_i = \omega_i + \sin( \varphi_{i+1} - \varphi_i) + \sin( \varphi_i
    - \varphi_{i-1}).
\end{equation}
The index $i$ denotes here the $i$-th phase in the chain. Note, that
the phase velocity is different for each oscillator.

The Thrust version for the coupled phase oscillator chain is very similar to
the Lorenz attractor example. Again, a zip iterator is used to pack the required
components and to process the resulting sequence with a single sweep of the
\code{for_each} algorithm. The only difference here is that values of
neighboring vector elements are needed. In order to access these values, we use Thrust's
permutation iterator, so that \code{operator()} of the system function object becomes
\begin{lstlisting}
thrust::for_each(
    thrust::make_zip_iterator(
        thrust::make_tuple(
            x.begin(),
            thrust::make_permutation_iterator( x.begin(), prev.begin() ),
            thrust::make_permutation_iterator( x.begin(), next.begin() ),
            omega.begin() , dxdt.begin() ) ),
    thrust::make_zip_iterator(
        thrust::make_tuple(
            x.end(),
            thrust::make_permutation_iterator( x.begin(), prev.end() ),
            thrust::make_permutation_iterator( x.begin(), next.end() ),
            omega.end(), dxdt.end() ) ),
    phase_oscillators_functor()
    );
\end{lstlisting}
Here, \code{prev} and \code{next} are vectors of type
\code{thrust::device_vector<size_t>} and hold the indices to the left and right
vector elements. The function object
\code{phase_oscillators_functor} implements \eqref{eq:phasesystem} similarly to the \code{lorenz_functor} above and is thus omitted for brevity.

The stencil operator in CMTL4 is a minimalistic matrix type.
Its application is expressed by a matrix-vector product that is assigned
to, or is used to either increment or decrement the vector:
\begin{lstlisting}
y = S * x;       y += S * x;      y -= S * x;
\end{lstlisting}
The user must provide a function object that applies the stencil on
the $i$-th element of a vector and its neighbors.
For the sake of performance the function object has to provide two methods:
one that is checking indices and to be applied near the beginning and the end
of the vector and the other without index checking.  For the considered
example, the function object is:
\begin{lstlisting}
struct stencil_kernel {
    static const int start = -1, end = 1;
    int n;

    stencil_kernel(int n) : n(n) {}

    template <typename Vector>
    __device__ __host__ double operator()(const Vector& v, int i) const {
	return sin(v[i+1] - v[i]) + sin(v[i] - v[i-1]);
    }

    template <typename Vector>
    __device__ __host__
    double outer_stencil(const Vector& v, int i, int offset= 0) const {
	double s1 = i > offset? sin(v[i] - v[i-1]) : sin(v[i]),
	       s2 = i+1 < n + offset? sin(v[i+1] - v[i]) : sin(v[i]);
	return s1 + s2;
    }
};
\end{lstlisting}
The parameter \code{offset} is needed when vector parts are cached so that the
addressing is shifted.
For the sake of backward (and forward) compatibility the non-portable keywords
\code{__device__} and \code{__host__} should be replaced by a macro that is
defined suitably for the according platform, e.g.~to an empty string on regular
compilers.  This makes the user code entirely platform-independent.

The stencil function object is passed as a template argument to the stencil matrix:
\begin{lstlisting}
typedef mtl::dense_vector<double>  state_type;

struct phase_oscillators {
    const state_type&  omega;
    mtl::matrix::stencil1D<stencil_kernel> S;

    phase_oscillators(const State& w) : omega(w), S(num_rows(w)) {}

    void operator()(const State &x, State &dxdt, double t) const {
	dxdt = S * x;
        dxdt += omega;
    }
};
\end{lstlisting}
The stencil matrix \code{S} in the system function above
uses shared memory to benefit from re-accessing vector entries
and to avoid non-coalesced memory accesses.

The VexCL version of the example is the most concise variant. The sum of
sines in \eqref{eq:phasesystem} is encoded using the
\code{VEX_STENCIL_OPERATOR} preprocessor macro. Its parameters are the name of the
resulting function object, the return type, the stencil width, the center, and the body string for
the generated OpenCL function encoding the required operation.  Once the stencil
operator is defined, \code{operator()} of the system function object is
implemented with a single line of code:
\begin{lstlisting}
typedef vex::vector<double> state_type;

struct phase_oscillators {
    const state_type &omega;
    phase_oscillators(const state_type &w) : omega(w) { }
    void operator()(const state_type &x, state_type &dxdt, double t) const {
        static VEX_STENCIL_OPERATOR(S, double, 3, 1,
                "return sin(X[1] - X[0]) + sin(X[0] - X[-1]);", omega.queue_list());
        dxdt = omega + S(x);
    }
};
\end{lstlisting}

The stencil operations are implemented in ViennaCL using the \code{shift()} operation,
which shifts the indices of a vector by a certain offset. 
New shifted values at the beginning or at the end of the vector are by default the same as the first or last entry of the vector, respectively,
which is just the required behavior for this example:
\begin{lstlisting}
typedef viennacl::vector<double> state_type;

struct phase_oscillators {
    const state_type &omega;
    phase_oscillators(const state_type &w) : omega(w) { }

    void operator()(const state_type &x, state_type &dxdt, double t) const {
      typedef viennacl::generator::vector<value_type> vec;

      viennacl::generator::custom_operation op;
      op.add( vec(dxdt) = vec(omega) + sin(shift(vec(x),  1) - vec(x))
                                      + sin(vec(x) - shift(vec(x), -1)) );
      op.execute();
    }
};
\end{lstlisting}

%
% DISORDERED LATTICES
%
\subsection{Disordered Hamiltonian Lattice}

The last example in our performance and usage study is a
nonlinear disordered Hamiltonian lattice \cite{mulansky_scaling_2012}. Its
equations of motion are governed by
\begin{equation}
\dot{q}_{i,j} = p_{i,j}, \quad \quad
\dot{p}_{i,j} = - \omega_{i,j}^2 q_{i,j} - \beta q_{i,j}^3 + \Delta_d q_{i,j}.
\label{eq:disordered_ham}
\end{equation}
Here, $\Delta_d q_{i,j}$ denotes the two-dimensional discrete Laplacian
$\Delta_d
q_{i,j}=q_{i+1,j}+q_{i-1,j}+q_{i,j+1}+q_{i,j-1}-4q_{i,j}$. Such
systems are widely used in theoretical physics to study phenomena
like Anderson localization \cite{Sheng-06} or thermalization \cite{FPUScholarpedia}.

An important property of \eqref{eq:disordered_ham} is its Hamiltonian
nature. It can be obtained from the Hamilton equations and energy as well as phase volume conservation during the time
evolution can be shown. To account for these properties, a special class of solvers
exists, namely symplectic solvers. Odeint implements three different
variants of such solvers, all being of the Runge-Kutta-Nystr\"om
type \cite{HairerGeometricNumericalIntegration2006,Leimkuhler-Reich-04}. The
implementation of these solvers requires only the second part
of \eqref{eq:disordered_ham} with $\dot{p}_{i,j}$ to be specified by
the user.

The natural choice for the implementation of \eqref{eq:disordered_ham} is a
sparse matrix-vector product. Since Thrust neither provides sparse matrix
types nor sparse matrix-vector products, Thrust was combined with the
CUSPARSE library in order to implement this example. CUSPARSE contains a set of
basic linear algebra subroutines used for handling sparse matrices and is
included in the CUDA Toolkit distribution together with the Thrust library
\cite{NvidiaCusparseManual}.

For better comparison, all libraries considered in our study use the hybrid ELL
format for storing the sparse matrix on GPUs, since it is one of the most
efficient formats for sparse matrices on these devices~\cite{BellGarland2008}.
The standard compressed sparse row format is used for CPU runs.  As the
construction of the sparse matrix for $- \omega_{i,j}^2 q_i + \Delta_d q_{i,j}$
is straight-forward, we only provide code for the system function object
interface.

The relevant code for the Thrust version of the system function object is
\begin{lstlisting}
typedef thrust::device_vector<double> state_type;

void operator()(const state_type &q , state_type &dp) const {
    static double one = 1;
    thrust::transform(q.begin(), q.end(), dp.begin(), scaled_pow3_functor(-beta) );

    cusparseDhybmv(handle, CUSPARSE_OPERATION_NON_TRANSPOSE,
            &one, descr, A, thrust::raw_pointer_cast(&q[0]), &one,
            thrust::raw_pointer_cast(&dp[0]) );
}
\end{lstlisting}
Here, \code{handle}, \code{descr}, and \code{A} are CUSPARSE data structures
holding the CUSPARSE context and sparse matrix data. The
\code{thrust::transform()} algorithm is used in line 5 to compute the
scaled third power of the input vector \code{q}. Lines 7--9 call the sparse
matrix-vector product kernel in CUSPARSE, where
\code{thrust::raw_pointer_cast()} is used to convert the thrust device vector
iterator to a raw device pointer.

The CMTL4 implementation reads:
\begin{lstlisting}
typedef mtl::dense_vector<double> state_type;

void operator()(const state_type& q, state_type& dp) {
    dp = A * q;
    dp -= beta * q * q * q;
}
\end{lstlisting}
Here \code{A} is an instance of a sparse matrix holding the discretization of
the linear combination $- \omega_{i,j}^2 q_i + \Delta_d q_{i,j}$. The
expression \code{q * q * q} computes the triple element-wise product of column
vector \code{q}.  Usually, products of column/row vectors among themselves are
often program errors and therefore not allowed in CMTL4.  Their use may be
enabled by defining an according macro during compilation.

The VexCL version employs the user-defined OpenCL function \code{pow3}, which
computes the third power of its argument and is used for the sake of best
performance:
\begin{lstlisting}
typedef vex::vector<double> state_type;
VEX_FUNCTION(pow3, value_type(value_type),  "return prm1 * prm1 * prm1;");

void operator()(const state_type &q, state_type &dp) const {
    dp = (-beta) * pow3(q) + A * q;
}
\end{lstlisting}

Similar to CMTL4, the ViennaCL version of the system function object is split into two parts:
first, the sparse matrix-vector product $Aq$ is computed; second, the
non-linear term $-\beta q^3$ is added to the result by means of a custom
operation:
\begin{lstlisting}
typedef viennacl::vector<double> state_type;

void operator()(const state_type &q, state_type &dp) const {
    typedef viennacl::generator::vector<value_type> vec;

    dp = viennacl::linalg::prod(m_A, q);
    viennacl::generator::custom_operation op;
    op.add( vec(dp) -= m_beta * element_prod(vec(q), element_prod(vec(q), vec(q))) );
    op.execute();
}
\end{lstlisting}

\section{Results} \label{sec:results}

We present results obtained from our numerical experiments in this section. The
complete source code for the experiments and the full set of results are freely
available in a GitHub repository\footnote{ \href{
https://github.com/ddemidov/gpgpu_with_modern_cpp } {
https://github.com/ddemidov/gpgpu\_with\_modern\_cpp } }.

All the libraries tested in this paper with the exception of CMTL4 and CUSPARSE
allow for the use of both CPU and GPU devices.  Thrust supports an OpenMP-based
execution on the CPU, which is enabled by a compilation switch, while OpenCL
libraries natively support CPUs provided that the respective runtime is
installed. OpenCL implementations from AMD and from Intel were used on the CPU and from AMD and NVIDIA on GPUs.  The
timings provided were obtained on a Gentoo Linux operating system for two GPUs,
namely an NVIDIA Tesla C2070 and an AMD Radeon HD~7970 (Tahiti), as well as for
an Intel Core i7 930 CPU.  All reported values are median values of execution
times taken for ten runs.

\begin{figure}
    \begin{center}
      \includegraphics[width=\textwidth]{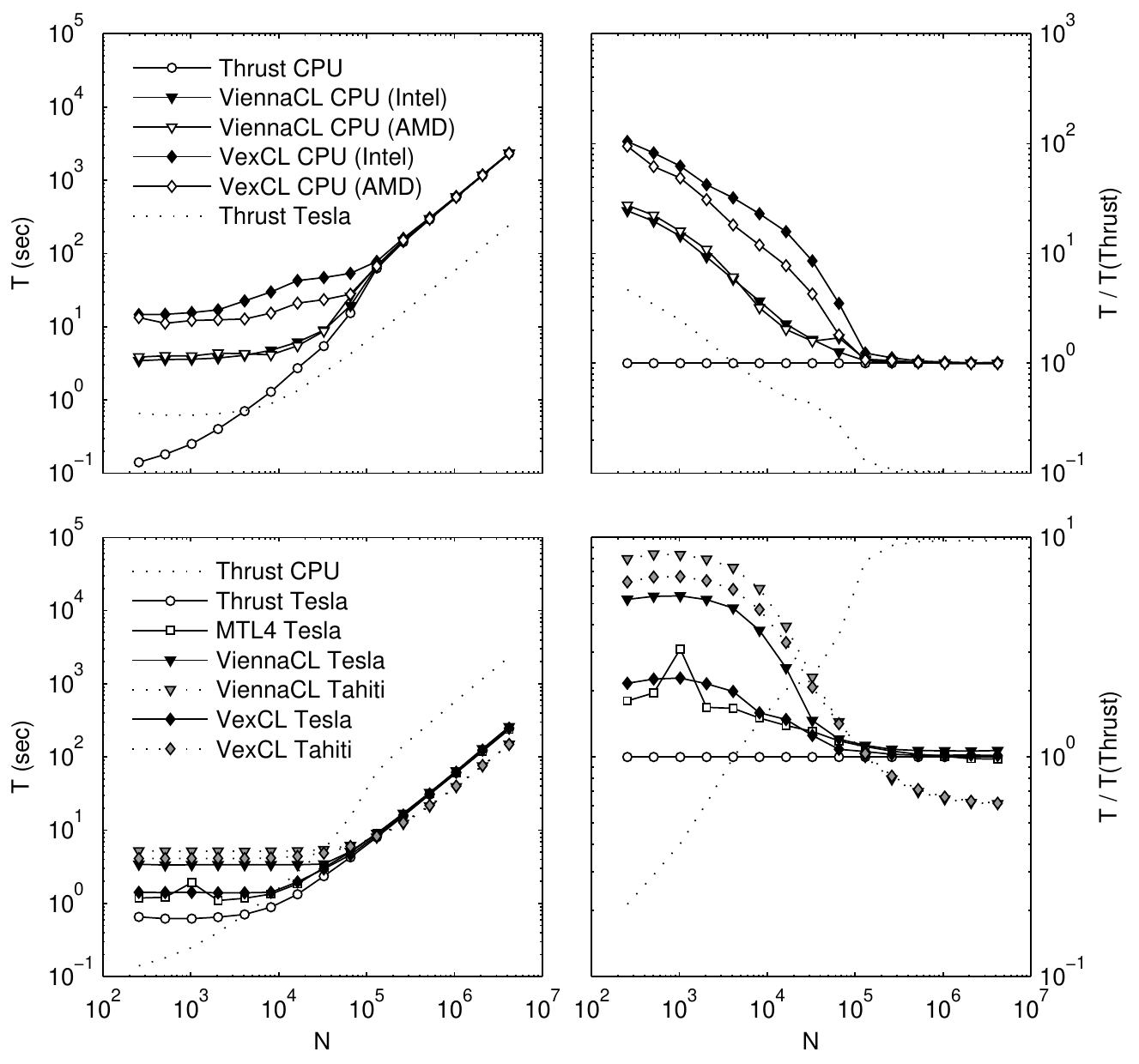}
    \end{center}
    \caption{Lorenz attractor ensemble results.}
    \label{fig:lorenz:perf}
\end{figure}

\begin{figure}
    \begin{center}
        \includegraphics[width=\textwidth]{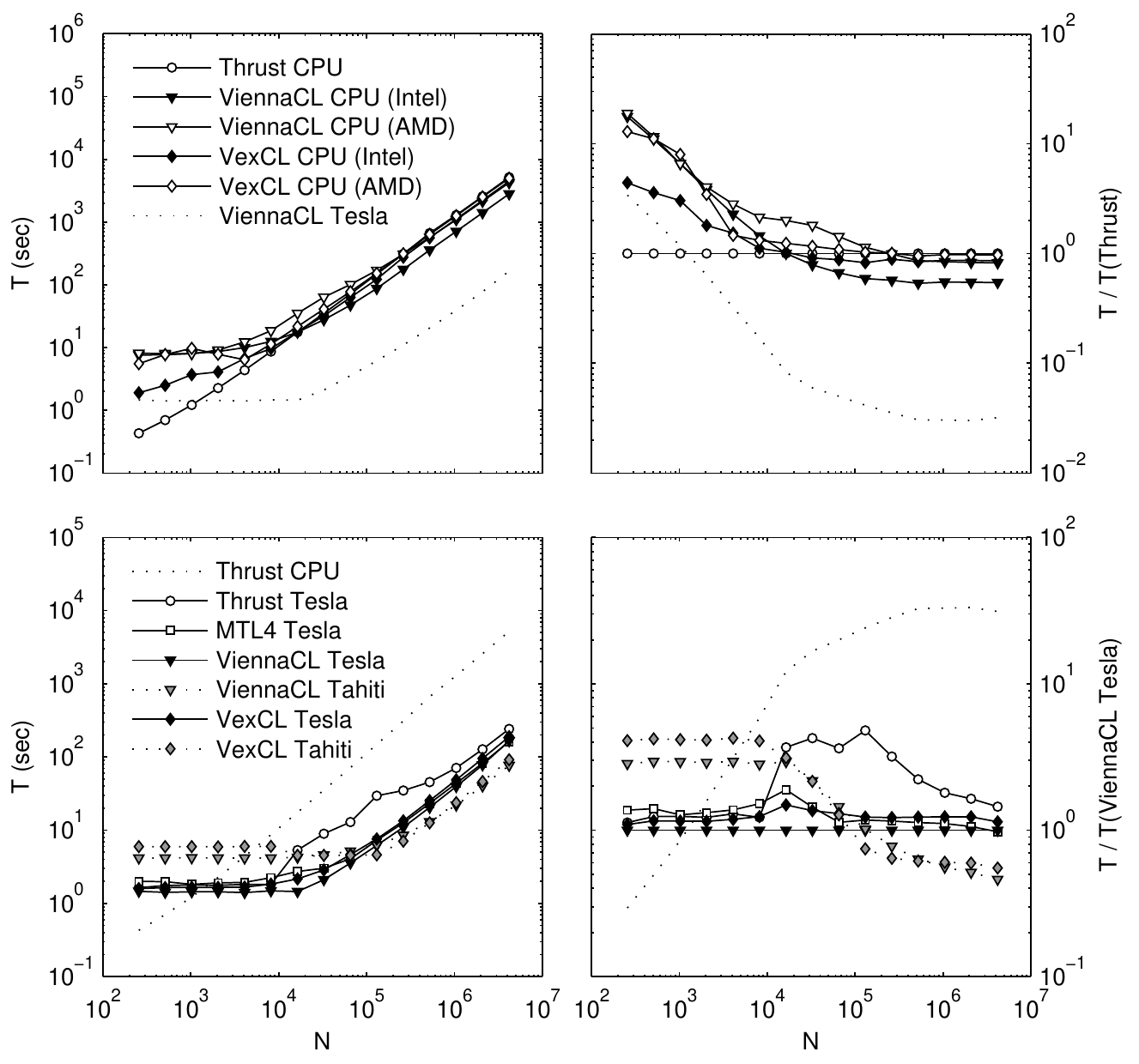}
    \end{center}
    \caption{Coupled phase oscillator chain results.}
    \label{fig:phase:perf}
\end{figure}

\begin{figure}
    \begin{center}
        \includegraphics[width=\textwidth]{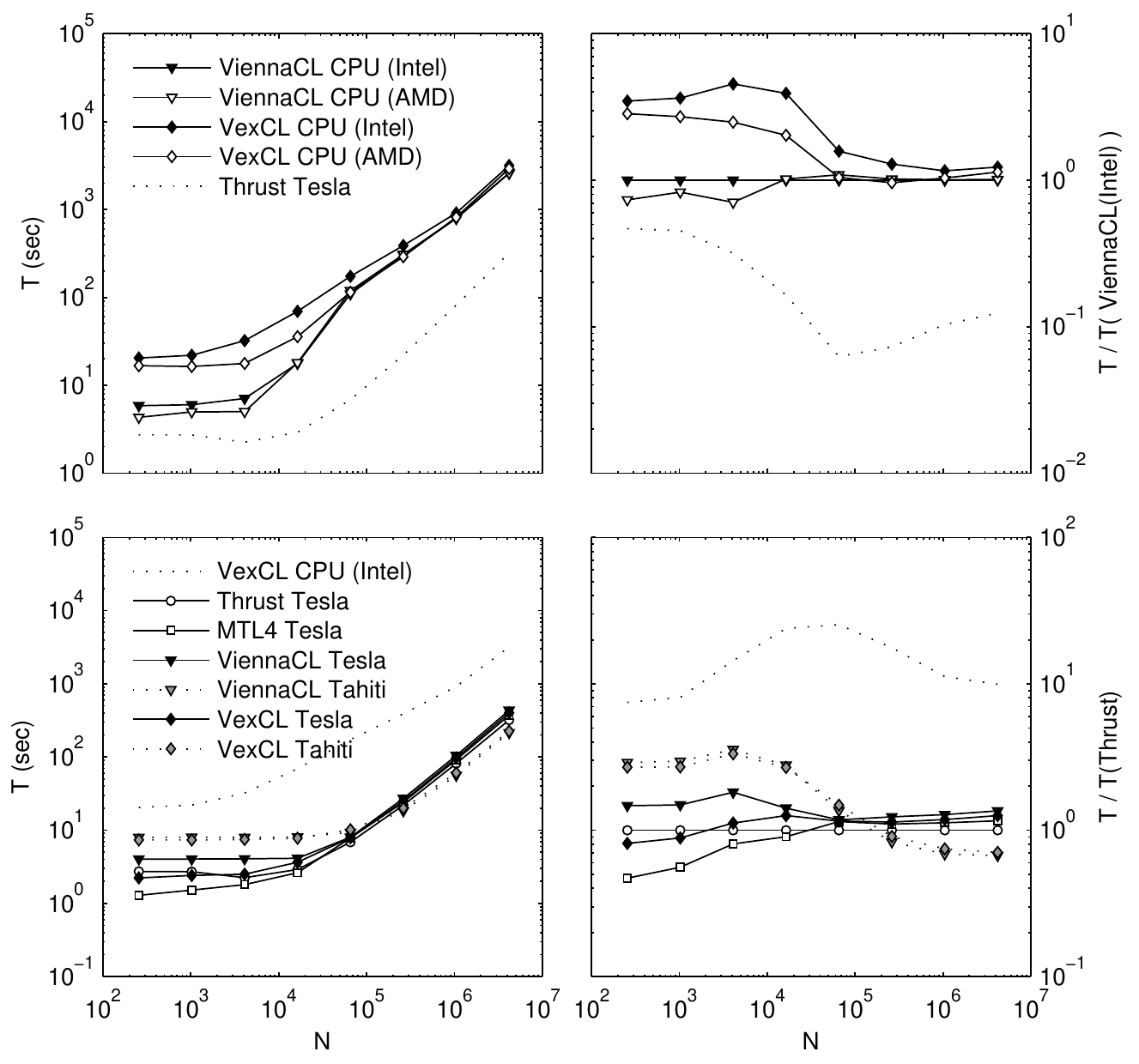}
    \end{center}
    \caption{Disordered Hamiltonian lattice results.}
    \label{fig:lattice:perf}
\end{figure}

Figures \ref{fig:lorenz:perf} through \ref{fig:lattice:perf} show performance
data for the three examples discussed in the previous section.  The top row in
each figure shows the performance obtained from CPU-based experiments, while
bottom rows shows GPU-based data. On the GPU plots the graphs for NVIDIA Tesla
and AMD Tahiti boards are correspondingly plotted with solid and dotted lines.
The plots on the left show absolute solution time over the size of the problem
being solved, while the plots on the right depict performances relative to the
Thrust version with two exceptions, where ViennaCL is selected as the reference
library. The first exception are GPU plots on \figref{fig:phase:perf}, where
Thrust library performs badly in case of the coupled phase oscillator chain
example. The other exception is \figref{fig:lattice:perf}, where the
combination of Thrust and CUSPARSE is not able to run on a CPU. 

Absolute execution times for the largest problem size for all of the considered
libraries are given in Table~\ref{tab:abstimes}. The table also provides
the achieved memory bandwidth in GB/sec and in fractions of the theoretical peak
for each of the compute devices.

\subsection{GPU-Performance}
In general, all our experiments show up to $10\times$ to $20\times$
acceleration when run on a GPU as compared to the CPU path. This is the
expected acceleration rate for the memory bandwidth bound examples that we
looked at. However, both CUDA and OpenCL programs show considerable overhead
at smaller problem sizes, thus requiring problems of sizes between $10^3$
and $10^5$ to see any significant acceleration on a GPU at all. The overhead
for OpenCL libraries is larger than that of CUDA programs, which is mostly due
to the additional kernel management logic required by the OpenCL runtime.

Performance-wise, all of the considered libraries are close to each other when
run on a GPU.  VexCL and ViennaCL are in general slower than CMTL4 and Thrust
by a few percent, which is usually negligible in practice.  Apparently, the
CUSPARSE implementation of the sparse matrix-vector product is more efficient
than that of the rest of the libraries, since it outperforms the competitors by
about 20-30\% in the disordered Hamiltonian lattice experiment.  The
implementation of the phase oscillator chain example for Thrust is rather
ineffective since it uses a permutation iterator requiring an additional
global vector for storing indices. The implementations of stencil operations 
in CMTL4 and ViennaCL are slightly more efficient than those of VexCL.

Moreover, the overhead of using high-level libraries is negligible compared
to the effort spent in getting familiar with the details of CUDA or OpenCL.
Thus, we have successfully countered productivity concerns for GPU computing
raised in the past~\cite{bordawekar:gpu-productivity}.

\subsection{CPU-Performance}

Thrust, VexCL, and ViennaCL show very similar performance on the CPU for
larger problem sizes. For smaller problems the difference between Thrust
and OpenCL-based libraries is more pronounced, since Thrust uses an OpenMP backend
which does not have any overhead such as OpenCL initialization and kernel
compilation.  The difference between the OpenCL
implementations of AMD and the Intel is negligible in most cases. The only
exception is the example of the chain of phase oscillators, where the
implementation by Intel outperforms the one of AMD by up to 50 percent.  This
might be explained by either a better implementation of trigonometric functions
in Intel's version, or the autovectorization feature of Intel's OpenCL SDK,
which transparently compiles OpenCL kernels to fully utilize the SIMD
processing on the underlying Intel CPU.

It has to be said that the overhead of OpenCL for small problem sizes is
tremendous, if not embarrassing, hence OpenCL cannot be considered to be a
competitive CPU programming model for a large area of applications in its
present state. 
A considerable reduction in kernel launch overhead for CPU-based kernel execution is required to make OpenCL more attractive on this target.

% GBytes transfered for each of the experiments:
%   Lorenz:      25625.0
%   Oscillator:  11875.0
%   Ham lattice: 38437.5
% Theoretical peaks used:
%   Tesla:       148  GB/sec (http://www.nvidia.com/object/personal-supercomputing.html)
%   Tahiti:      264  GB/sec (http://www.amd.com/us/products/desktop/graphics/7000/7970/Pages/radeon-7970.aspx#3)
%   Core i7-930: 25.6 GB/sec (http://ark.intel.com/products/41447/Intel-Core-i7-930-Processor-8M-Cache-2_80-GHz-4_80-GTs-Intel-QPI)  
\begin{table}
    \begin{small}
    \caption{Absolute run times (sec) and achieved throughput (GB/sec and
    percentage of theoretical peak) for the largest problem size.}
    \label{tab:abstimes}
    \begin{tabular}{|l|rrrrrr|}
        \hline
        & \multicolumn{2}{|c|}{Lorenz attractor}
        & \multicolumn{2}{|c|}{Phase oscillators}
        & \multicolumn{2}{|c|}{Hamiltonian lattice} \\
        \cline{2-7}
        & \multicolumn{1}{c}{Time} & \multicolumn{1}{|c|}{T-put}
        & \multicolumn{1}{c}{Time} & \multicolumn{1}{|c|}{T-put}
        & \multicolumn{1}{c}{Time} & \multicolumn{1}{|c|}{T-put} \\
        \hline
        \multicolumn{7}{|c|}{NVIDIA Tesla C2070} \\
        \hline
        Thrust           &   242.78 & 105 (71\%) &   240.87 &  49 (33\%) &  319.60 & 120 (81\%) \\
        CMTL4            &   237.91 & 108 (73\%) &   161.96 &  73 (50\%) &  370.31 & 104 (70\%) \\
        VexCL            &   246.58 & 104 (70\%) &   189.38 &  63 (42\%) &  401.39 &  96 (65\%) \\
        ViennaCL         &   259.85 &  99 (66\%) &   166.20 &  71 (48\%) &  433.50 &  89 (60\%) \\
        \hline
        \multicolumn{7}{|c|}{AMD Radeon HD 7970 (Tahiti)} \\
        \hline
        VexCL            &   149.49 & 171 (65\%) &    91.60 & 130 (49\%) &  225.41 & 170 (65\%) \\
        ViennaCL         &   148.69 & 172 (65\%) &    76.55 & 155 (59\%) &  214.87 & 179 (68\%) \\
        \hline
        \multicolumn{7}{|c|}{Intel Core i7 930} \\
        \hline
        Thrust           & 2~336.14 &  11 (43\%) & 5~182.55 &   2 ( 9\%) & \multicolumn{2}{c|}{N/A} \\
        VexCL    (AMD)   & 2~329.00 &  11 (43\%) & 5~011.66 &   2 ( 9\%) & 2~934.99 &  13 (51\%) \\
        VexCL    (Intel) & 2~372.70 &  11 (42\%) & 4~463.24 &   3 (10\%) & 3~171.74 &  12 (47\%) \\
        ViennaCL (AMD)   & 2~322.78 &  11 (43\%) & 4~246.24 &   3 (11\%) & 2~608.80 &  15 (58\%) \\
        ViennaCL (Intel) & 2~322.39 &  11 (43\%) & 2~815.23 &   4 (16\%) & 2~580.47 &  15 (58\%) \\
        \hline
    \end{tabular}
    \end{small}
\end{table}

\begin{figure}
    \begin{center}
        \subfigure[
        Lorenz attractor ensemble.
        ]{\includegraphics[width=0.32\textwidth]{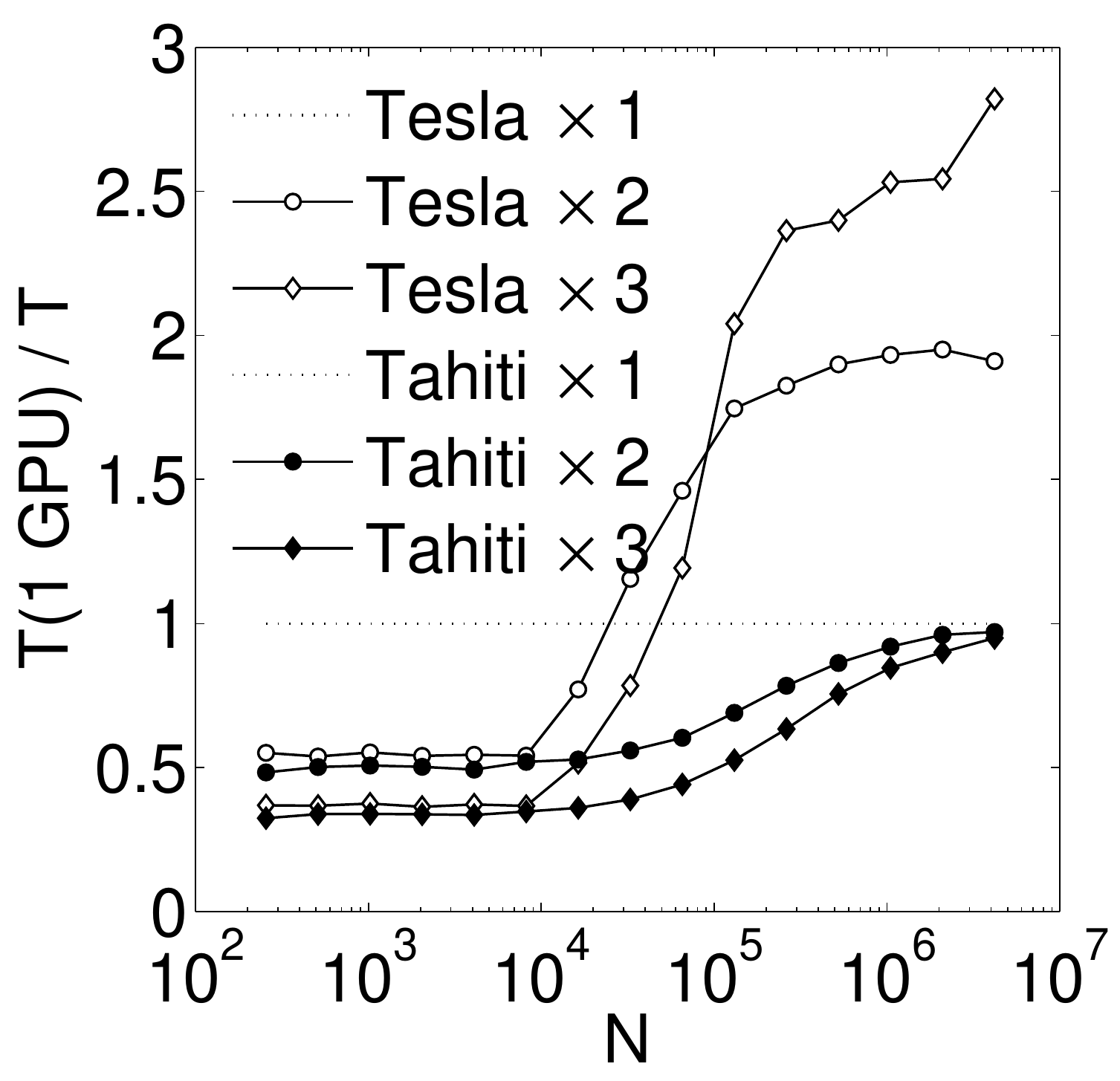}}$\;$
        \subfigure[
        Coupled phase oscillator chain.
        ]{\includegraphics[width=0.32\textwidth]{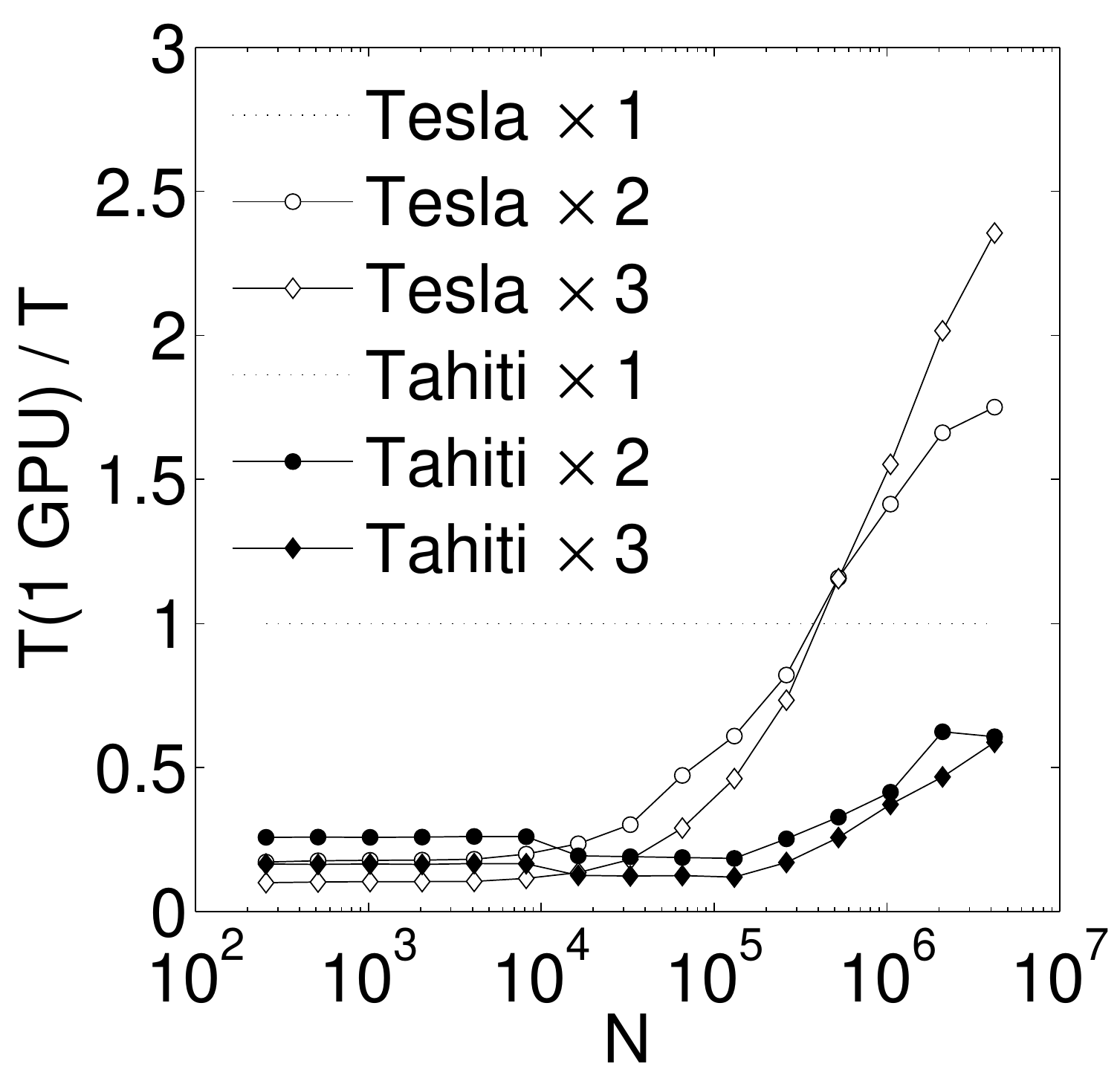}}$\;$
        \subfigure[
        Disordered Hamiltonian lattice.
        ]{\includegraphics[width=0.32\textwidth]{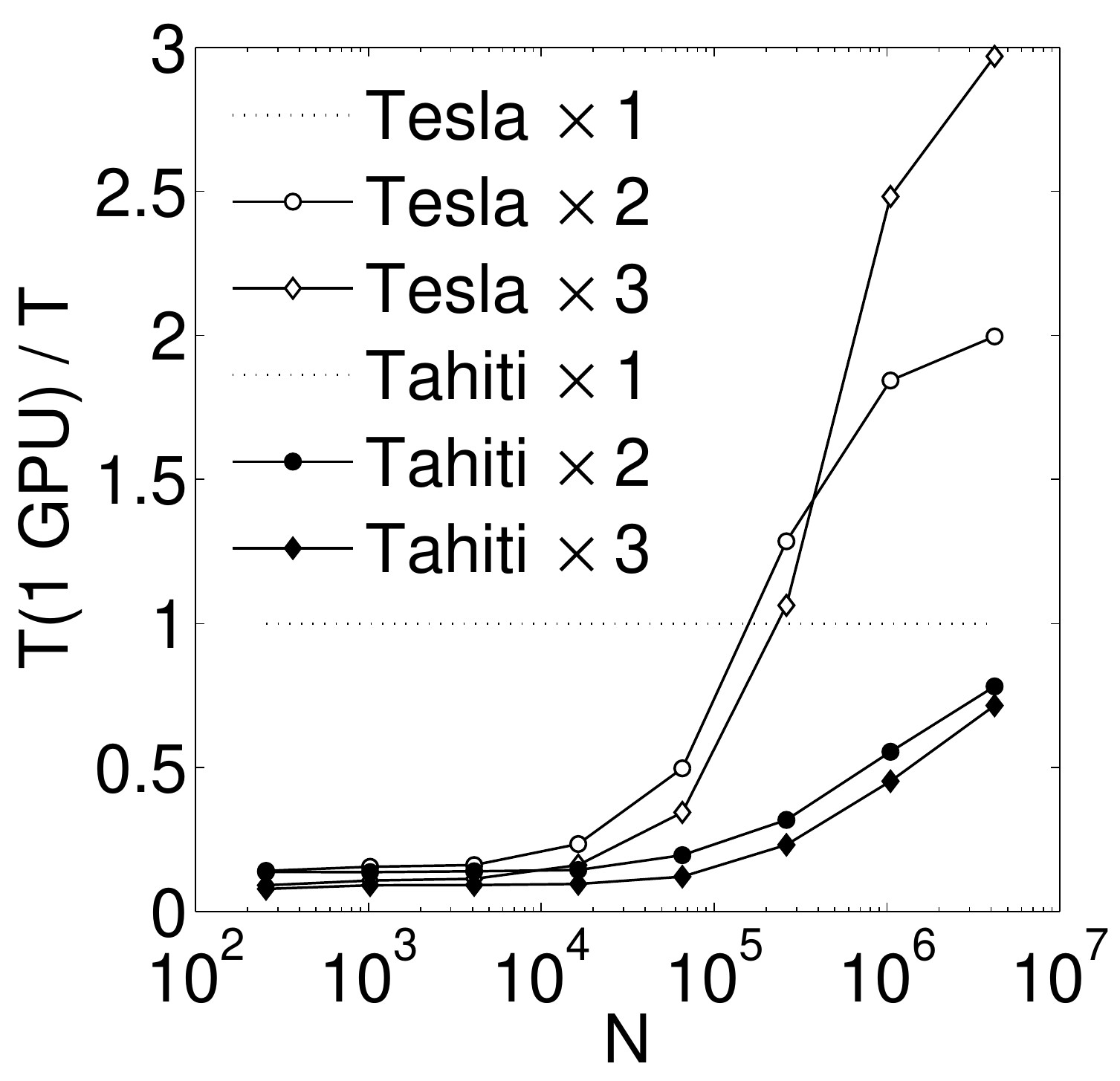}}
    \end{center}
    \caption{VexCL scaling with multigpu computation.}
    \label{fig:scaling}
\end{figure}

Finally, results for multi-GPU usage as provided by VexCL in a transparent way
are considered. \figref{fig:scaling} shows scaling results for up to three
GPUs. It can be seen that a notable speed-up for several Tesla GPUs over a
single card is only obtained for problem sizes larger than $10^6$.  It seems
that AMD's OpenCL implementation does not work very well with multiple GPUs
employed. Still, the combined memory of several GPUs allows to solve proportionally
larger problems on the same system.

%
% CONCLUSION
%
\section{Conclusion}

Performance-wise, there is almost no difference between various platforms and
libraries when those are run on the same hardware for large problem sizes.
As we have shown, various
computational problems may be solved effectively in terms of both human and
machine time with the help of modern high-level libraries. Hence, the
differences in the programming interfaces of the libraries are more likely to
determine the choice of a particular library for a specific application rather than raw performance.

The focus of Thrust is more on providing low-level primitives with an
interface very close to the \Cpp Standard Template Library. Special purpose
functionality is available via separate libraries such as CUSPARSE and
can be integrated without a lot of effort.  The rest of the libraries
we looked at demonstrated that they are able to provide a more
convenient interface for a scientific programmer than a direct
implementation in CUDA or OpenCL.  CMTL4 and VexCL have a richer set of
element-wise vector operations and allow for the shortest
implementations in the context of the ODEs considered in this work.
ViennaCL requires two to three additional lines of code in each of the examples due to the use of the kernel generator facility.
Still, this extra effort is acceptable considering that the library's focus is on sparse
linear systems solvers, which are, however, beyond the scope of this
paper.

Regarding a comparison of CUDA versus OpenCL, the main difference observed in
this work is the wider range of hardware supported by OpenCL.  Although the performance
obtained via CUDA is a few percent better than that of OpenCL on the overall,
the differences are mostly too small in order to make a decision in favor of
CUDA based on performance only.  Moreover, the slight performance advantage of
CUDA can still turn into a disadvantage when taking the larger set of hardware
supporting OpenCL into consideration.

% Another aspect that has not been studied in this work is the ability to
% generate OpenCL kernels optimized for the problem at hand at runtime. This
% allows, for example, to generate optimized kernels for a certain set of
% parameters supplied, eliminating any otherwise spurious reads from global
% memory.  An in-depth study of such an approach is, however, left for future
% work.

\section{Acknowledgments}

This work has been partially supported by the Russian Foundation for Basic
Research (RFBR) grant No 12-07-0007 and the Austrian Science Fund (FWF), grant
P23598.  We also would like to thank Gradient JSC\footnote{ \href{
http://www.gradient-geo.com/en }{ http://www.gradient-geo.com/en } } for the
kindly provided AMD hardware.

\bibliographystyle{siam}
% \bibliography{ref}

\end{document}